\documentclass[11pt]{article}
 \pdfoutput=1
\usepackage{amssymb}
\usepackage{xcolor}
\usepackage[pdftex]{graphicx}
\usepackage{epsfig}
\usepackage{cite}
\newcommand{\AmS}{{\protect\the\textfont
  A\kern-.1667em\lower.5ex\hbox{M}\kern-.125emS}}
\newcommand{\ba}{\begin{array}}
\newcommand{\ea}{\end{array}}

\def\beq{\begin{equation}}
\def\eeq{\end{equation}}

\def\beq{\begin{equation}}   
\def\eeq{\end{equation}}

\def\bea{\begin{eqnarray}}
\def\eea{\end{eqnarray}}

\begin{document}
\begin{titlepage}

\begin{flushright}
 IFIC/22-18, FTUV-22-06-24
\end{flushright}

\vskip 1.5cm

\begin{center}
{\huge{\bf
Stringy signals from large-angle correlations\\ 
\vskip 0.5cm
in the cosmic microwave background?}}
\end{center}

\vspace{2.0cm}

\begin{center}

{\bf Miguel-Angel Sanchis-Lozano$^{\ast}$}
\vspace{1.5cm}\\

\it Instituto de F\'{\i}sica
Corpuscular (IFIC) and Departamento de F\'{\i}sica Te\'orica \\
{\it CSIC-University of Valencia, Dr. Moliner 50, E-46100 Burjassot, Spain}

\end{center}

\vspace{0.5cm}

\begin{abstract}
We interpret 
the lack of large-angle temperature correlations 
and the apparent even-odd parity imbalance, observed in 
the cosmic microwave background by COBE, WMAP and {\em Planck} satellite 
missions, as a possible stringy signal ultimately stemming
from a composite inflaton field (e.g. a fermionic condensate). Based on causality arguments and a Fourier analysis of the angular two-point correlation function, {\em two} infrared cutoffs $k_{\rm min}^{\rm even,odd}$ are introduced in the CMB power spectrum associated, respectively, with periodic and antiperiodic boundary conditions of the fermionic constituents (echoing the Neveu-Schwarz-Ramond model in superstring theory), without resorting to any particular model.
\end{abstract}

\begin{center}


\end{center}
\vskip 2.5cm

\begin{small}
\noindent 
$^\ast$E-mail address: Miguel.Angel.Sanchis@ific.uv.es
\end{small}

\end{titlepage}

\bigskip

\section{Introduction}

Since its discovery in 1964 by Penzias and Wilson, the study of the 
cosmic microwave background (CMB) has provided a wealth of information on the
early universe and its evolution, even going back in time  
closer to the Big Bang than the
recombination phase when it was released. Indeed, to explain the outstanding homogeneity of the CMB temperature across the sky, an inflationary phase has been put forward to solve the so-called horizon problem, together with the flatness issue and
$\lq\lq$unwanted" relics, like unseen magnetic monopoles \cite{Kolb(1994)}.

Although there are alternative models (see e.g. \cite{DiValentino(2021)}, \cite{Brandenberger(2009)}), 
inflation has currently become the commonly 
accepted paradigm to explain cosmological evolution because of a series of convincing reasons of which we
just highlight one: observations of the CMB temperatures across the sky show that the 
universe is approximately uniform and homogeneous at scales much larger than
the particle horizon at decoupling time, and a phase of exponential growth can account for this.

In principle, all what is needed for a successful inflation is a scalar field satisfying the slow-roll conditions and a graceful exit. Actually the inflaton needs not
be a fundamental (nor single) scalar field but an effective, e.g. composite, field 
(see Ref.\cite{Samart(2022)} for a review) stemming from
more fundamental interacting fields (like a fermionic condensate 
\footnote{Historically, the interpretation of scalars
as fermionic bound states, e.g. in 
the Nambu-Jona-Lasinio model, 
dates back to the sixties in the past century, to address strong interaction issues
\cite{Nambu(1961)},\cite{Gross(1974)}). The possibility of a composite Higgs boson is also contemplated \cite{Gunion(1989)}.}). 
Through the process of inflation, initial small
quantum fluctuations of the underlying 
inflationary field when exiting the Hubble horizon
would have formed the seeds for the matter
distribution and subsequent growth to 
structures seen in today's universe, showing an overall isotropy and homogeneity at large scales. Alternatively, in a linearly expanding cosmology, like the so-called $R_h=ct$ universe (for an exhaustive review of this model 
see \cite{Meliab(2020)}), the quantum 
fluctuations of the primordial field driving the expansion 
turned into (semi-)classical fluctuations once exiting the Planck domain at about the Planck time \cite{Melia(2021)}.

On the other hand, in spite of the many successes of the Standard Cosmological Model ($\Lambda$CDM), anomalies and tensions have emerged from a variety of astrophysical and cosmological
observations (see e.g. \cite{Perivolaropoulos(2021)},\cite{DiValentino(2021)},\cite{Sunny(2020)}). In particular, 
the observed but unexpected
lack of large-angle angular correlations (related to the missing power at low multipoles),
together with the odd-parity dominance in the two-point correlation function of the CMB
were examined in detail in Ref.\cite{Sanchis(2022)}. To this end, an infrared cutoff was set in the primordial CMB power spectrum 
following the original work of \cite{Melia(2018)}. At the same time, 
odd-parity dominance was taken into account by weighting the different multipole contributions to angular correlations. 
Needless to say, such a parity imbalance questions the large-scale isotropy of the observable 
universe and hence the Cosmological Principle. 

Here we shall perform a similar analysis but introducing two infrared cutoffs to the 
CMB power spectrum ($k_{\rm min}^{\rm even}$, $k_{\rm min}^{\rm odd}$), 
affecting differently 
even- and odd-parity multipole contributions in the
best fit of the
two-point correlation function $C(\theta)$. Firstly, the 
aim of using two infrared cutoffs, rather than one,
is to simultaneously achieve both goals (reproducing missing large-angle correlations and
odd-parity dominance) with fewer fitting parameters, while suggesting 
an interesting and novel physical interpretation. Indeed, 
large-angle correlations should provide details on the very 
first stages of the universe 
maybe hinting new physics
at unexplored large scales. This possibility stands as
the main motivation of this paper.

In order to get the best fit of $C(\theta)$ to {\em Planck} 2018 data, 
we find that the ratio $k_{\rm min}^{\rm even}/k_{\rm min}^{\rm odd}$ should be
numerically close to 2. Later, we will provide a tentative theoretical 
explanation for this, 
 based on an assumed $\lq\lq$stringy" \footnote{In this paper
{\em stringy} refers generically to properties 
associated with periodic and antiperiodic boundary conditions like those found 
in superstring theory \cite{Zwiebach(2004)}, and not, in principle, to cosmic strings
understood as one-dimensional structures of false vacuum \cite{Kolb(1994)}.}
behaviour of the fundamental (fermionic) 
field(s) driving inflation in the early universe.

\section{Analysis of angular correlations in the CMB}
All three COBE, WMAP and {\em Planck} satellite missions have observed that the temperature angular distribution of the CMB is
remarkably homogeneous across the sky, with anisotropies of order 1 part in $10^5$. A powerful test of these
fluctuations relies on the two-point (and higher)
angular correlation function $C(\theta)$, defined as the ensemble product
of the temperature differences with respect to the average temperature, from all pairs of directions
in the sky defined by unitary vectors $\vec{n}_1$ and $\vec{n}_2$:
\begin{equation}\label{eq:CTT}
C(\theta)=\langle \frac{\delta T(\vec{n}_1)}{T}\frac{\delta T(\vec{n}_2)}{T} \rangle\;,
\end{equation}
where $\theta \in [0,\pi]$ is the angle defined by the scalar product $\vec{n}_1 \cdot \vec{n}_2$.

Assuming azimuthal symmetry, $C(\theta)$ can be expanded using the
Legendre polynomials as:
\begin{equation}\label{eq:C2}
C(\theta)=\frac{1}{4\pi}\sum_{\ell=2}^{\infty}(2\ell+1)\ C_{\ell}\ P_{\ell}(\cos(\theta)\;,
\end{equation}
where the $C_{\ell}$ multipole coefficients encode the information with astrophysical/cosmological significance. In practice, the sum starts at $\ell=2$ and ends at a given $\ell_{\rm max}$, dictated by the
resolution of the data (in our case we set $\ell=400$ as an  upper limit). The first two terms are excluded because (i) the monopole ($\ell=0$)
is simply the average temperature over the whole sky and plays no role in the
correlations, other than a global scale shift; and (ii) the dipole ($\ell=1$) is
greatly affected by Earth's motion, creating an anisotropy dominating over the intrinsic
cosmological dipole signal, being usually removed
from the multipole analysis.

Under some simplifying assumptions \cite{Mukhanov(2005)}, the coefficients $C_{\ell}$ in Eq.(\ref{eq:C2}) can be evaluated according to
the following expression 
\begin{equation}\label{eq:Cell}
C_{\ell}\ =\ N\ \int_{0}^{\infty}dk\ k^{n_s-1}\ j_{\ell}^2(kr_{d})\;,
\end{equation}
where $N$ is a normalization constant, $n_s \approx 1$ is the spectral index and
$j_{\ell}(kr_{d})$ denotes the spherical Bessel function
of order $\ell$, whose argument involves the comoving distance $r_d$ from the last scattering surface  (LSS) to us, and the wavenumber $k$ for each mode in 
the CMB power spectrum 
which varies, in principle, from zero to infinity \cite{Mukhanov(2005)}.

\subsection{Single infrared cutoff in the CMB power spectrum}

As already commented, large-angle temperature correlations in the CMB provide information on the earliest stages of the
primitive Universe, well before recombination and the subsequent formation of the cosmic structure \cite{Mukhanov(2005)}. In this regard,  
the $C(\theta)$ function defined in Eq.(\ref{eq:C2}) 
was found close to zero above $60^{\circ}-70^{\circ}$ 
from all three COBE, WMAP and {\em Planck} data, constituting
one of the anomalies between standard cosmology
and observations \cite{Perivolaropoulos(2021)}. Indeed, 
the suppression of large-angle correlations together with 
the existence of a 
downward tail at large angles ($\gtrsim 150^\circ$)
were unexpected in standard cosmology, given that inflation is supposed to provide 
the required number of e-folds ($\gtrsim 60)$
to solve both the horizon and flatness problems, thereby 
bringing angular coverage across the full sky.

In order to mitigate this tension, the authors of \cite{Melia(2018)}
introduced an infrared cutoff in the CMB power spectrum, leading
to a lower limit $k_{\rm min}$ in the integral:
\begin{equation}\label{eq:Cellcutoff}
C_{\ell}\ =\ N\ \int_{k_{\rm min}}^{\infty}dk\ k^{n_s-1}\ j_{\ell}^2(kr_{d})\;.
\end{equation}
The normalization constant $N$ and 
$k_{\rm min}$ 
are obtained from a global fit to the whole angular correlation function.

\begin{figure}
\begin{center}
\includegraphics[width=8.6cm]{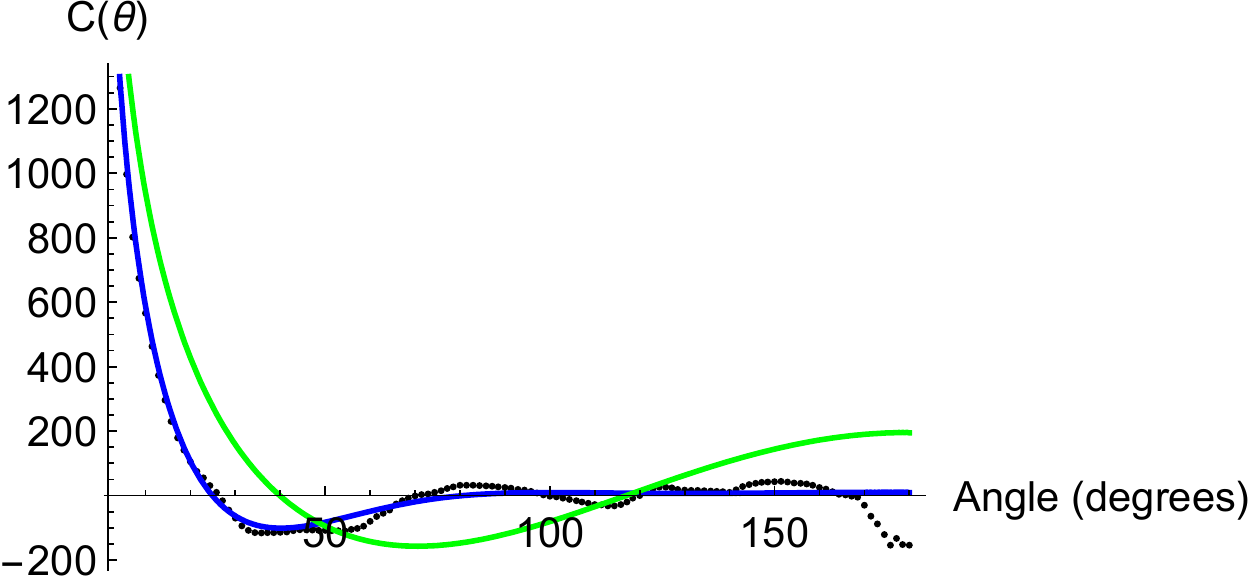}
\includegraphics[width=8.6cm]{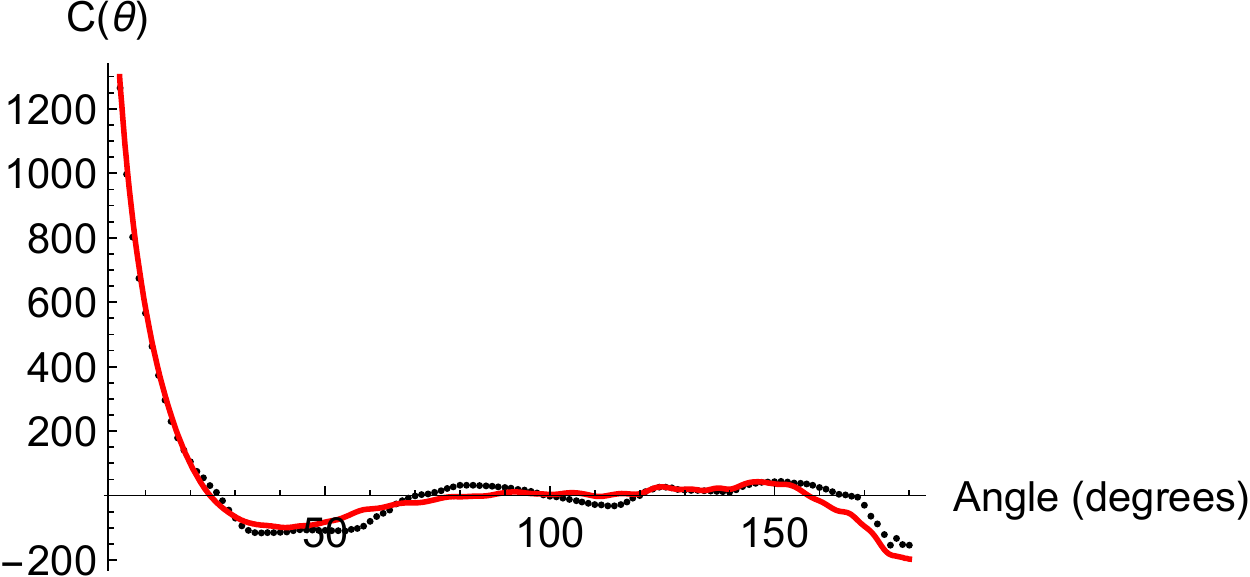}
\caption{Two-point correlation function $C(\theta)$) (solid curves)
obtained from the fit to {\em Planck} 2018 data (\textbf{a}) Left panel: $\Lambda$CDM
prediction (in green) clearly deviating from data in almost the whole angular range;
blue curve: $u_{\rm min}=4.5$ without imposing odd-dominance requirement. 
(\textbf{b}) Right panel: Red curve for $u_{\rm min}^{\rm even}=4.5$ with odd-parity dominance from \cite{Sanchis(2022)}.}
\end{center}
\label{fig1}
\end{figure}

A theoretical motivation to this truncation 
of $k$ modes entering the integral of Eq.(\ref{eq:Cellcutoff}), relies
on the existence of a maximum correlation length obtained from a
causality requirement at decoupling time $t_d$
\cite{Melia(2012)},\cite{Melia(2018)}: 
\begin{equation}\label{eq:lambdamax}
    \lambda_{\rm max}=2\pi R_h\;,
\end{equation}
where $R_h$ stands for the Hubble radius at that time. Actually
the Hubble radius $R_h$ not necessarily is a true
horizon, but should provide an order of magnitude
estimate of the maximum correlation length.
In fact $\lambda_{\rm max}= \alpha 2\pi R_h$, where $\alpha \leq 1$
depends on the cosmological model, e.g. $\alpha \approx 0.5$
for the $\Lambda$CDM \cite{Melia(2013)}. 

On the other hand, the comoving wavenumber $k_{\rm min}$ associated with $\lambda_{\rm max}$ reads
\begin{equation}
k_{\rm min}=\frac{2\pi a(t)}{\lambda_{\rm max}}\;,
\end{equation}
which can be interpreted as
the first quantum fluctuation in the underlying
primordial field driving the early universe expansion
either (i) having crossed the Hubble horizon
once inflation started, or (ii) 
having emerged out of the Planck domain \cite{Melia(2021)} 
should inflation have never happened, as 
in a $R_h=ct$ universe \cite {Meliab(2020)}. 

Next, changing the integration variable in Equation~(\ref{eq:Cell}) from $k$ to 
the dimensionless variable $u\equiv kr_d$ 
and setting $n_s=1$ for simplicity, one gets \begin{equation}\label{eq:Cumin}
C_{\ell}\ =\ N\ \int_{u_{\rm min}}^{\infty}\ du\ \frac{j_{\ell}^{\,2}(u)}{u}\;.
\end{equation}
Let us point out that only those $C_{\ell}$
coefficients with $\ell \lesssim 10$ are actually affected by the
lower cutoff $u_{\rm min} = k_{\rm min}r_d \neq 0$ in the above integral.

In Ref.\cite{Sanchis(2022)} the numerical interval $u_{\rm min}=4.5\pm 0.5$
was obtained from a best fit to the {\it Planck} 2018 dataset \cite{Planck(2018)}. 
In the left panel of Figure 1 we  
plot $C(\theta)$ corresponding to 
the $\Lambda$CDM prediction ($u_{\rm min}=0$), together with the best fit to 
the same {\it Planck} data as done in 
 \cite{Melia(2018)} with $u_{\rm min}=4.5$ 
without breaking even-odd parity balance. 
One can see at once that standard cosmology fails to fit the
data, while the latter curve 
indeed yields almost zero correlations above $\simeq 70^\circ$, but fails to 
reproduce the observed downward tail at $\sim 180^\circ$.
Finally, in the right panel of Figure 1 we reproduce the 
excellent fit ($\chi^2/{\rm d.o.f.} \approx 1$)
done in \cite{Sanchis(2022)} 
to the same dataset requiring 
odd-parity dominance.

On the other hand, the maximum correlation angle $\theta_{\rm max}$ can be roughly estimated as
\begin{equation}\label{eq:angmax}
    \theta_{\rm max}\simeq \frac{\lambda_{\rm max}}{R_d}\ \to\ \theta_{\rm max}\simeq \frac{2\pi}{u_{\rm min}}\;,
\end{equation}
where $R_d=a(t_d)r_d$ stands for the proper distance from the LSS to us, and $u_{\rm min}=k_{\rm min}r_d$.

\begin{figure}
\begin{center}
\includegraphics[width=10.5cm]{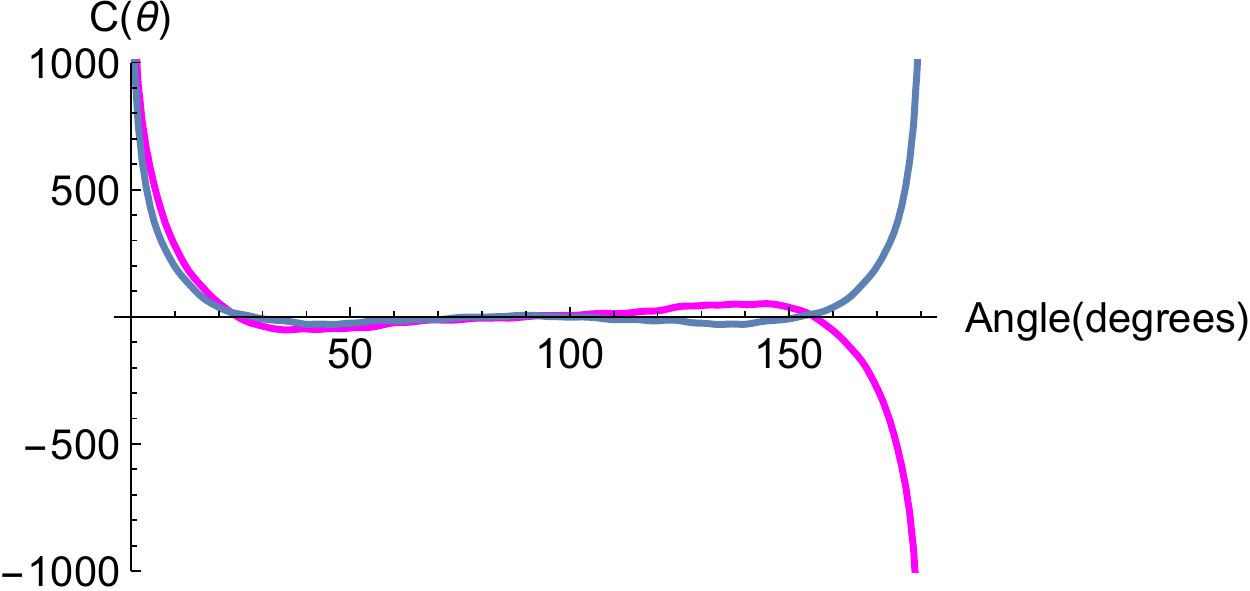}
\caption{The $C_{\rm even}(\theta)$ (in blue) 
and $C_{\rm odd}(\theta)$ (in magenta) contributions
to $C(\theta)$ in Eq.~(\ref{eq:Csplit}), setting $u_{\rm min}=4.5$. Notice 
the delicate balance between odd and even $\ell$-multipoles (even at high-$\ell$) in order to yield zero correlations above $\approx 60^{\circ}-70^{\circ}$. Any slight deviation from this balance would yield a downwards (upwards) tail at $\simeq 180^{\circ}$ for
odd (even) dominance of Legendre polynomials.}
\end{center}
\label{fig2}
\end{figure}

For later use, let us regroup the even and odd $\ell$-multipoles in Eq.(\ref{eq:C2}), namely
\begin{eqnarray}\label{eq:Csplit}
C(\theta)&=& C_{\rm even}(\theta)+C_{\rm odd}(\theta)= \nonumber \\ &&\frac{1}{4\pi}\sum_{\ell_{\rm even}}(2\ell+1)\ C_{\ell}\ P_{\ell}(\cos(\theta)+\frac{1}{4\pi}\sum_{\ell_{\rm  odd}}(2\ell+1)\ C_{\ell}\ P_{\ell}(\cos(\theta)
\end{eqnarray}

To better understand the behaviour of the $C(\theta)$ curve, 
we plot in Figure 2 separately the above even and odd parity
pieces ($C_{\rm even}(\theta)$ and $C_{\rm odd}(\theta)$),  
as a function of $\theta$ for $\ell \in [2,400]$.
Because of the oscillatory behaviour of the Legendre polynomials, both contributions add positively at small and middle angles, while a delicate balance is needed in order to get zero correlation at larger angles. This balance is indeed achieved 
\cite{Sanchis(2022)} when using a single cutoff $k_{\rm min}$
in the evaluation of the $C_{\ell}$ coefficients in Eq.(\ref{eq:Cumin}).

However, if {\em two} distinct infrared cutoffs apply 
differently to even and odd $\ell$-modes, such a delicate balance will be likely broken. Then odd or even parity dominance can be obtained {\em naturally}. In the next section we examine in depth this issue of paramount importance in this work.

\subsection{Two infrared cutoffs in the CMB power spectrum}

In Ref.\cite{Sanchis(2022)}, the fit to {\em Planck} 2018 data was optimized
using a $k_{\rm min}$ together with the requirement of parity imbalance by weighting adequately the odd and even terms in Eq.(\ref{eq:C2}).
We shall now carry out a similar analysis but introducing two
infrared cutoffs (rather than one) in the 
CMB power spectrum, namely, $k_{\rm min}^{\rm even}$ and $k_{\rm min}^{\rm odd}$,
affecting differently
the even- and odd-parity coefficients. 

The goals of this new analysis are to provide: 
(i) a common explanation of both missing large-angle correlations and
odd-parity dominance with fewer fitting parameters; and (ii) a 
physical interpretation behind.
First, the numerical values of $k_{\rm min}^{\rm even}$ and $k_{\rm min}^{\rm odd}$
were heuristically determined
from a fit of $C(\theta)$ to data, checking its consistency 
with odd-dominance by means of a parity statistic 
to be discussed in section 3.1. Depending on which, 
$k_{\rm min}^{\rm even}$ or $k_{\rm min}^{\rm odd}$ is larger, odd or even dominance
can be achieved.

\begin{figure}
\begin{center}
\includegraphics[width=10.5cm]{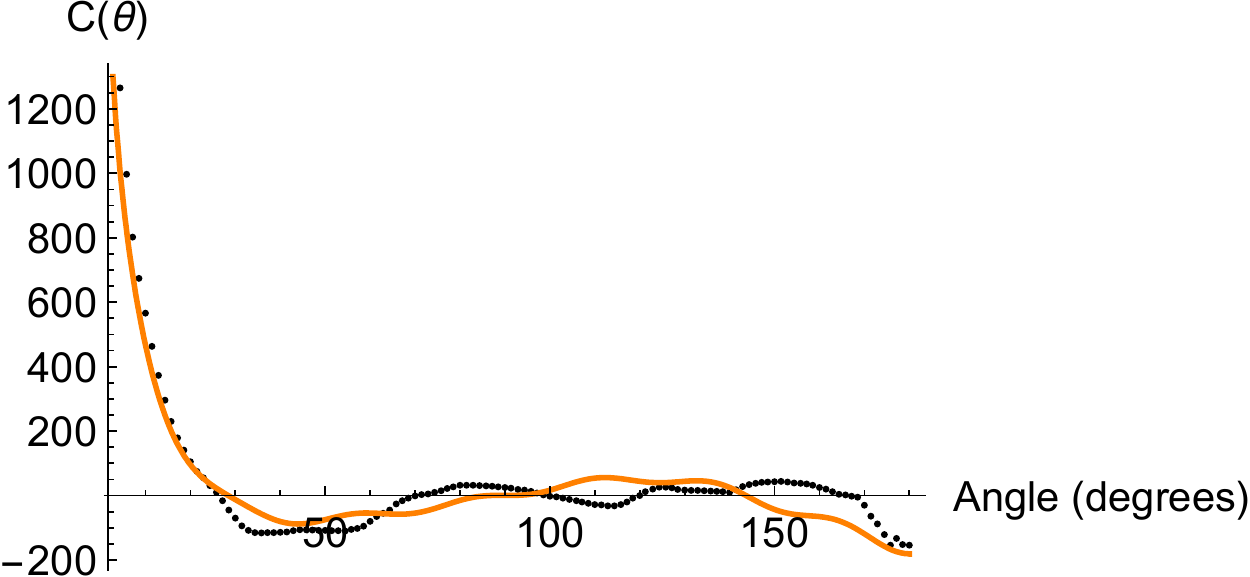}
\caption{Two-point correlation function $C(\theta)$ (orange solid curve)
obtained from the fit to {\em Planck} 2018 data for $u_{\rm min}^{\rm even}=5.34$ and  $u_{\rm min}^{\rm even}= u_{\rm min}^{\rm even}/2=2.67$. The fit is not so good as in the right panel of Figure 1, but the number of fitting parameters is 
quite smaller and, above all, provides
a theoretical foundation that explains simultaneously the lack of
large-angle correlations and the observed odd-parity dominance
leading to a downward tail near $180^{\circ}$.}
\end{center}
\label{fig3}
\end{figure}   

On the one hand, the introduction of two infrared cutoffs
in the angular power spectrum can be merely seen as a phenoenological approach
to jointly account for missing large-angle correlations and the
downwards tail in the $C(\theta)$ curve. Seen in this way, just one additional fitting parameter is required 
with respect to the analysis of \cite{Melia(2018)}. On the other hand, however, 
we claim in this work that the ratio of both infrared cutoffs 
becomes $\lq\lq$fixed by theory'', as a consequence of an extra (fermionic) spin degree of freedom
due to the assumed fermionic nature of the underlying inflating field(s). Hence
no additional parameter is actually introduced in the fits.

Therefore we compute separately the $C_{\ell}$ coefficients of
the even and odd $\ell$-multipoles according to 
\begin{equation}\label{eq:Cellcutoffs}
C_{\ell_{\rm even/odd}}\ =\ N\ \int_{u_{\rm min}^{\rm even/odd}}^{\infty}du\ \frac{j_{\ell}^2(u)}{u}\;,
\end{equation}
equivalent to Eq.(\ref{eq:Cumin}) for a single $k_{\rm min}$
without a distinction of parity. Now the integral lower limits are: $u_{\rm min}^{\rm even,odd}=k_{\rm min}^{\rm even,odd}r_d$, respectively, where $r_d$ denotes 
again the distance from the LSS to us. 

In Figure 3 we plot the fit of $C(\theta)$ to {\em Planck} 2018 data under
the assumption of 
two different 
lower cutoffs $u_{\rm min}^{\rm even,odd}$ in the integral 
of Eq.(\ref{eq:Cellcutoffs}). Admittedly, the new fit
looks a bit worse than in the right panel of Figure 1. However
notice that the number of free parameters now is smaller and, above all, 
one can simultaneously accomodate the lack of
large-angle correlations and the apparent odd-dominance showing up as 
a downward tail at $\gtrsim 150^{\circ}$, without resorting to
any further fine-tuning of the $C_{\ell}$ coefficients. 

Furthermore, a suggestive interpretation emerges as 
the ratio $k_{\rm min}^{\rm even}$ to $k_{\rm min}^{\rm  odd}$ turns out to be numerically
near 2. Had we tried the opposite situation
$k_{\rm min}^{\rm even}<k_{\rm min}^{\rm  odd}$, the $C(\theta)$ curve
would have shown an upward tail, contrary to
observational data.
The mathematical reason why
the condition $k_{\rm min}^{\rm even}>k_{\rm min}^{\rm  odd}$ goes in the $\lq\lq$right" direction is that the $C_{\ell}$ coefficients (for $\ell \lesssim 10$) 
decrease for $\ell_{\rm odd}$ to a lower extent than 
for $\ell_{\rm even}$ (as compared to no lower cutoff at all), thereby favoring odd-parity dominance, in accordance with observations.

Next, notice the following relations between Legendre polynomials and the square of cosine functions  
with entire and half-entire ($2\pi$) periods (these relations 
can also be formulated in terms of Chebyshev polynomials 
\cite{Abramowitz(1970)}), called to 
play a fundamental role in our later development:
\begin{eqnarray}\label{eq:Legendre}
    P_1(\cos{\theta}) & = & -1+2\cos^2{(\theta/2)} \nonumber \\
P_2(\cos{\theta}) & = & -0.5+1.5\cos^2{(\theta)} \nonumber  \\
P_3(\cos{\theta}) & = & -1+0.75\cos^2{(\theta/2)}+1.25\cos^2{(3\theta/2)} \nonumber \\
P_4(\cos{\theta}) & = & -0.7184+0.6249\cos^2{(\theta)}+1.0937\cos^2{(2\theta)} \nonumber \\
P_5(\cos{\theta}) & = & -1+0.4687\cos^2{(\theta/2)}+0.5469\cos^2{(3\theta/2)}+
0.9844\cos^2{(5\theta/2)}\ ;\ {\rm etc} 
\end{eqnarray}
Higher order Legrende polynomials replicate the same pattern: besides a constant, even and odd polynomials   
either contain $\cos^2{[n\theta]}$ or $\cos^2{[(n+1/2)\theta]}$ terms, respectively. Note that this pattern is just mathematical for the moment, but 
below will be put in correspondence 
with the Fourier expansion of inflaton field constituents, using a one-dimensional toy-model.

Next, inspired on very general grounds 
by superstring theory (and obviously without entering at all in its complexity and intrincacies \cite{Zwiebach(2004)}), 
and invoking a similar
causality argument as employed for the single $k_{\rm min}$
case, we address now the
possibility that any constituent (fermionic) field $\psi(\varphi)$
(for simplicity no more variables or indices are explicitly written)
of a composite inflaton
\footnote{The possibility that a fermionic field, minimally or non-minimally coupled with the gravitational field, drives the inflation, without resorting to a composite inflaton, has also been studied (see \cite{Grams(2014)} and references therein).}
satisfies either the periodic or
antiperiodic boundary condition:
\begin{eqnarray}\label{eq:periantiperi}
\psi(\varphi+2\pi) & = & \psi(\varphi) \\  
\psi(\varphi+2\pi) & = & -\psi(\varphi)\ \to\ \psi(\varphi+4\pi)  = \psi(\varphi)
\end{eqnarray}

The angular Fourier expansion of $\psi(\varphi)$ 
for the periodic condition reads:
\begin{equation}\label{eq:periodic}
\psi(\varphi)=\sum_{n \in {\cal Z}}\alpha_n\ e^{in\varphi}
\end{equation}
and for a real function
\begin{equation}\label{eq:cos}
\psi(\varphi)=2\sum_{n \in {\cal Z^+}}{\rm Re}\ \alpha_n\ e^{in\varphi}\;,
\end{equation}
so that only $\cos{(n\varphi)}$ terms appear in the Fourier expansion.

Let us define now a correlation function as in \cite{Copi(2018)} 
\begin{equation}
    \int_{0}^{2\pi}\psi(\varphi)\ \psi(\varphi+\Delta \varphi)\ \frac{d\varphi}{2\pi}\;.
\end{equation}
For random Gaussian Fourier coefficients, if we define $\theta=\Delta \varphi/2$ one finds
\begin{equation}
    C(\Delta \varphi)=2\sum_{n \in {\cal Z^+}}C_n \cos{(n\Delta\varphi)}\ \to\  
    C(\theta)=4\sum_{n \in {\cal Z}^+}C_n \cos^2{(n\theta)}-2\;,
\end{equation}
with $C_n=\langle\alpha_n \alpha_n^{\ast}\rangle$ and
$\theta \in [0,\pi]$
to be identified with the angle appearing as the argument
of the two-point correlation function $C(\theta)$.

For the antiperiodic conditions, the Fourier expansion reads
\begin{equation}\label{eq:antiperiodic}
    \psi(\varphi)=\sum_{n \in {\cal Z}^+ + 1/2}\alpha_n\ e^{in\varphi}\;,
\end{equation}
which guarantees that it changes sign when $\varphi \to \varphi+2\pi$. Therefore
\begin{equation}
    C(\theta)=4\sum_{n \in {\cal Z^+}+1/2}C_n \cos^2{(n\theta)}-2\;.
\end{equation}

The above results obtained from this toy-model 
suggest the assignment of even- and odd-parity 
$C_{\rm even}(\theta)$ and $C_{\rm odd}(\theta)$ pieces 
in Eq.(\ref{eq:Csplit}), to 
the periodicity and antiperiodicity conditions 
given in Eqs.(\ref{eq:periodic}-\ref{eq:antiperiodic}), 
respectively, somewhat recalling the
well-known Ramond and Neveu-Schwarz sectors 
in superstring theory \cite{Zwiebach(2004)}.
Hereafter we shall assume that the above
conclusions apply to the fundamental field(s) driving the expansion of the
early universe.

\section{Physical interpretation}

In the following we will consider, without entering into details, the 
inflaton as an 
effective scalar field $\Phi$ made out of two
constituent fermionic fields ($\Phi=\bar{\Psi}\Psi$) 
driving the expansion of the early universe. 
The causality condition yielding $\lambda_{\rm max}$ used for a single infrared cutoff, is assumed now to
split into two conditions applied independently to
the periodic and the antiperiodic conditions  
in virtue 
of the spin degrees of freedom of the constituent fermionic
fields of the effective inflaton.

Hence, in analogy to Eq.(\ref{eq:lambdamax}), we shall now consider two maximum correlation lengths for even and odd parity multipoles, 
denoted 
as $\lambda_{\rm max}^{\rm even}$ and $\lambda_{\rm max}^{\rm odd}$, satisfying
\begin{equation}\label{eq:lambdas}
    \lambda_{\rm max}^{\rm even}=2\pi R_h,\ \ \lambda_{\rm max}^{\rm odd}=4\pi R_h.
\end{equation}
Then two comoving wavenumbers $k_{\rm max}^{\rm even}$ and $k_{\rm max}^{\rm odd}$
can be defined from those length scales:
\begin{equation}\label{eq:ks}
k_{\rm min}^{\rm odd}=2\pi a(t_d)/\lambda_{\rm max}^{\rm odd},\ \ \  k_{\rm min}^{\rm even}=2\pi a(t_d)/\lambda_{\rm max}^{\rm even}\;,
\end{equation}
corresponding to odd and even parity modes, respectively; $a(t_d)$ denotes
the scale factor at decoupling time. Let us remark that physical
momenta are given by $k/a(t)$, but the ratio
$k_{\rm min}^{\rm even}/k_{\rm min}^{\rm  odd}$ remains frozen after exiting the 
horizon. Note also   
that the numerical values of the lower cutoffs are
obtained from a fit to correlation data, and 
not from Eq.(\ref{eq:lambdas}). Indeed 
what really matters in this work is 
that their ratio is equal (or close) to 2.

On the other hand, two relevant angles can be defined following Eq.(\ref{eq:angmax}) for a single $k_{\rm min}$:
\begin{equation}\label{eq:ang2}
     \theta_{\rm even}\ \simeq \frac{2\pi}{u_{\rm min}^{\rm even}}\ \simeq\  65^{\circ}\ \ ,\ \ 
     \theta_{\rm odd}\ \simeq\ \frac{2\pi}{u_{\rm min}^{\rm odd}}\ \simeq\ 130^{\circ}\;. 
\end{equation}
defining three different angular regions 
along the $0^{\circ}-180^{\circ}$ range where the
$C(\theta)$ function exhibits a different behaviour, 
as can be easily appreciated in Figures 1 and 3.

Now, from our discussion in the previous section on 
the Fourier expansion 
of the two-point angular correlation function, we will take into account that 
the $C_{\rm even}(\theta)$ and $C_{\rm odd}(\theta)$ pieces of Eq.(\ref{eq:Csplit})
are associated
to terms of the type $\cos^2{[n\theta]}$
and $\cos^2{[(n+1/2)\theta]}$, respectively, as shown in 
Eqs.(\ref{eq:Legendre}). Therefore, the lower limits for even ond odd multipoles
in the integral of 
Eq.(\ref{eq:Cellcutoffs}) can be related to the
infrared cutoffs $k_{\rm min}^{\rm even,odd}$ 
($u_{\rm min}^{\rm even,odd}$), respectively. 

Moreover, from  the fit to the {\em Planck} 2018 data shown in Figure 3, the 
 ratio $u_{\rm min}^{\rm even}/u_{\rm min}^{\rm  odd}$
 turns out to be quite close to 2. This so far phenomenological fact
 will be understood further by considering the 
periodicity and antiperiodicity conditionsapplied to 
the maximum correlation lengths:
\begin{equation}
\frac{u_{\rm min}^{\rm even}}{u_{\rm min}^{\rm odd}}= \frac{k_{\rm min}^{\rm even}}{k_{\rm min}^{\rm odd}} = 
\frac{\lambda_{\rm max}^{\rm odd}}{\lambda_{\rm max}^{\rm even}}=2\;,
\end{equation}
as implied by Eq.(\ref{eq:lambdas}).
In the the framework of the $R_h=ct$ universe, this relation corresponds to two different exit times of the Planck regime: first the $k_{\rm min}^{\rm odd}$ mode and later the $k_{\rm min}^{\rm even}$ 
 mode. In an inflationary scenario, it implies again two different times 
 in a similar order but exiting the Hubble horizon. Needless to say, the above
 ratio 2 is expected to be valid only approximately, but we we have exactly fixed it to this value in our numerical analysis.

\begin{figure}
\begin{center}
\includegraphics[width=7.0cm]{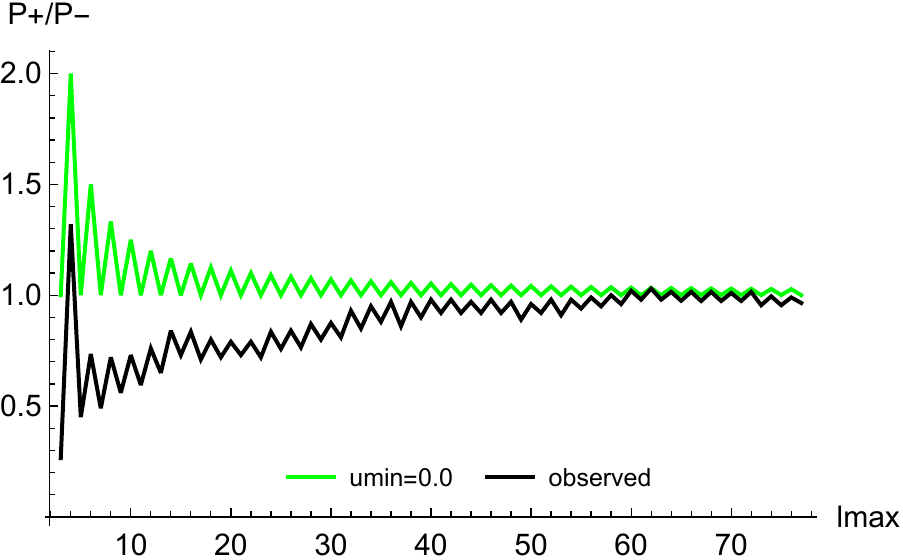}
\includegraphics[width=7.0cm]{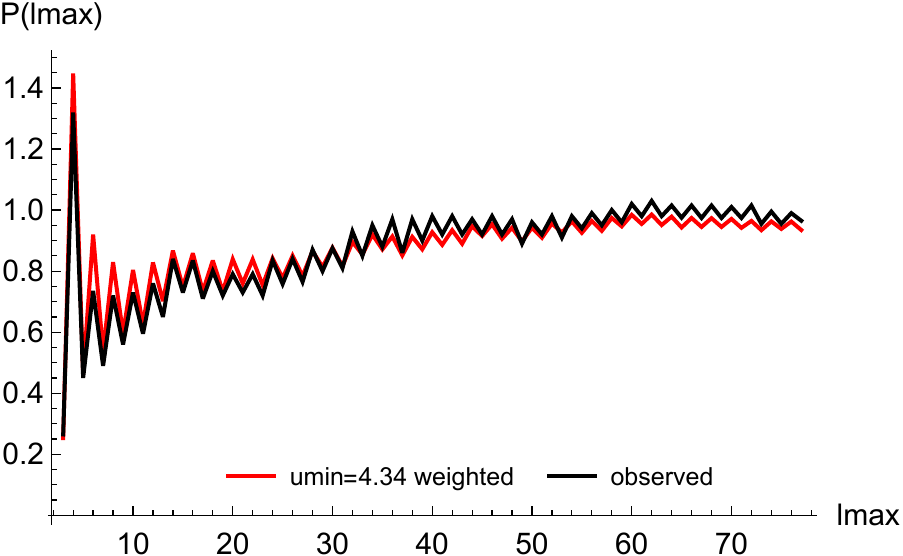}
\includegraphics[width=7.0cm]{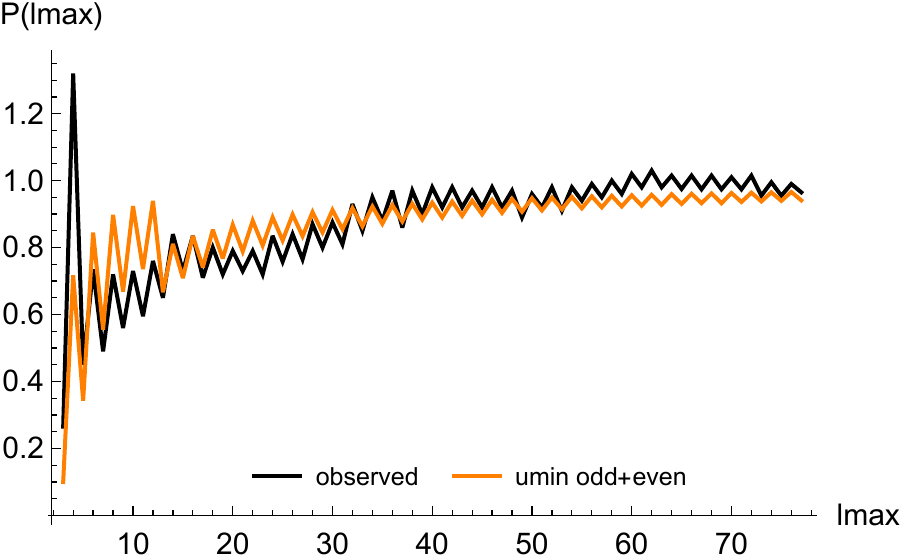}
\caption{$P(\ell_{\rm max})$ statistic as a function of $\ell_{\rm max}$  compared to the {\it Planck} 2018 data for
(\textbf{a}) Upper left panel: $u_{\rm min}=0$ and even-odd parity balance 
($\Lambda$CDM).
(\textbf{b}) Upper right panel: $u_{\rm min}=4.5$ and weighted $C_{\ell}$ to get odd-parity dominance ($\chi^2/{\rm d.o.f.} \approx 1$). 
(\textbf{b}) Lower panel: $u_{\rm min}^{\rm even}=5.34$ and $u_{\rm min}^{\rm odd}=2.67$ ($\chi^2/{\rm d.o.f.} \approx 2$).}
\end{center}
\label{fig4}
\end{figure}   

\newpage

\subsection{Parity statistic study}

let us now address the deviation from the even-odd parity balance 
employing the parity statistic \cite{Panda(2021)} as done in
\cite{Sanchis(2022)}
\begin{equation}
P(\ell_{\rm max})=\frac{P^+(\ell_{\rm max})}{P^-(\ell_{\rm max})}\ \ ;\ \ 
P^{\pm}(\ell_{\rm max})=\sum_{\ell=2}^{\ell_{\rm max}}\ 
\gamma_{\ell}^{\pm}\ \frac{\ell(\ell+1)}{2 \pi}\ C_{\ell}\;,
\end{equation}
with the projectors defined as $\gamma_{\ell}^+=\cos^2{(\ell\pi/2)}$ and
$\gamma_{\ell}^-=\sin^2{(\ell\pi/2)}$. Assuming that $\ell(\ell+1)\ C_{\ell} \sim \rm{constant}$
is satisfied at low $\ell$, $P^{\pm}$ can clearly be considered as a measurement of the degree
of parity asymmetry. Any deviation 
from unity of this statistic points to an even-odd parity imbalance: below unity, it implies odd-parity dominance and vice versa. Below we apply this statistic to compare different
fits with one or two infrared cutoffs.

In Figure 4 we plot $P(\ell_{\rm max})$ as a function of $\ell_{\rm max}$ corresponding to (a) upper left panel:
$u_{\rm min}=0$ ($\Lambda$CDM); (b) upper right panel: 
our previous result for $u_{\rm min}=4.5$ and odd-parity dominance \cite{Sanchis(2022)}; (c) lower panel: the new analysis 
performed in this paper assuming
$u_{\rm min}^{\rm even}=2u_{\rm min}^{\rm  odd}\simeq 5.34$. 
As an aside, notice that the mean value of both $u_{\rm min}^{\rm even,odd}$
remains within the previous interval $u_{\rm min}=4.5\pm 0.5$. 
From inspection of these plots it becomes apparent that the fit using  
$k_{\rm min}^{\rm odd,even}$ pair somewhat worsens
with respect
to the single $k_{\rm min} \neq 0$ case: the reduced $\chi^2/{\rm d.o.f.}$ increases
from almost unity to about twice. Nonetheless, as already emphasized, 
the main interest of providing a common explanation to both
missing large-angle correlations and odd-dominance, remains.

\section{Discussion and conclusions}

As claimed in Ref.\cite{Melia(2018)}, the lack of large-angle correlations observed
in the CMB angular distribution from WMAP and {\em Planck} 2018 data can 
be accounted for  
by introducing an infrared cutoff $k_{\rm min}$ to the CMB power
spectrum, thereby 
implying a lower cutoff $u_{\min}$ in the computation of all (odd and even)
coefficients $C_{\ell}$ of the multipole expansion in Eq.(\ref{eq:C2}). 
Furthermore, the apparent odd-parity dominance manifesting as 
a downward tail at large angles ($\theta \gtrsim 150^{\circ}$) 
can also be accomodated in the $C(\theta)$ plot
thanks to some extra weights mainly affecting 
the low-multipole coefficients $C_{\ell},\ (\ell \lesssim 10)$, leading 
at the same time to 
an excellent fit of the parity statistics 
$P(\ell_{\rm max})$ \cite{Sanchis(2022)}.

No obvious theoretical connection between such
an infrared cutoff $k_{\rm min}$ 
(i.e., lack of large-angle correlations) 
and parity imbalance has been
found so far to our knowledge, but stressed their
phenomenological compatibility 
\cite{Sanchis(2022)},\cite{Kim(2012)},\cite{Schwarz(2016)}. 
In this paper, however, we have put forward a possible
relationship between both anomalies by introducing to the CMB power spectrum 
two infrared cutoffs $k_{\rm min}^{\rm even}$ and $k_{\rm min}^{\rm  odd}$,
whose ratio turns out to be close to 2 from a fit to the 
{\em Planck} 2018 dataset.
In this way, we are able: (i) to reduce the number of fitting 
parameters with respect to \cite{Sanchis(2022)}, while reproducing the 
downward tail at large angles in the fit; (ii) to provide a theoretical
connection (without actually resorting to any particular model) 
between both observations, setting  
the ratio $k_{\rm min}^{\rm even}/k_{\rm min}^{\rm  odd}$  exactly equal
to 2. 

Thus, new fits of $C(\theta)$ and the parity statistic $P(\ell_{\rm max})$ 
have been 
performed using the same {\em Planck} 2018 dataset, for the double  
infrared cutoff case. Since the number of fitting parameters 
is now considerably smaller 
than in \cite{Sanchis(2022)}, 
the goodness of the fits (determined by their $\chi^2/{\rm d.o.f.}$) 
somewhat worsens. 
However, let us emphasize the added value of our approach 
due to a common and suggestive explanation of both anomalies: 
missing correlations
above $\approx 70^{\circ}$   
and a downward tail at $\gtrsim 150^{\circ}$.

On the basis 
of a Fourier analysis using
a toy-model, we have associated even and odd multipoles of 
$C(\theta)$ 
to periodic and antiperiodic boundary conditions satisfied  
by underlying (fermionic) fields of a composite inflaton, or 
a fermionic inflaton itself. Thus, the choice for the ratio
$k_{\rm min}^{\rm even}/k_{\rm min}^{\rm odd}$ = 2, becomes supported
by theoretical arguments based on spin, 
beyond the initial heuristic approach,  
while the fits remain statistically acceptable.

As is well-known, spin degrees of freedom play 
a crucial role in elementary particle physics, e.g., predicting the gyromagnetic ratio $g$ for the self-spinning electron to be about two times bigger than the value for an orbiting electron \cite{Thomson(2013)}. Needless to say, the famous muon $g-2$ deviation from zero stands as an important
test of new physics, currently under close scrutiny \cite{pdg(2020)}. 
The idea of searching for new physics and phenomena either at particle colliders and/or looking at the sky, in a complementary way, is certainly not new, being dark matter a good example of it.

Of course, explanations alternative to ours 
for missing large-angle correlations together with odd-parity 
dominance are possible: a statistical fluke of data at large angles, 
contamination effects, 
cosmic variance, or other underlying theoretical reasons.

Finally, let us point out that future very-high precision 
measurements of CMB polarization
\cite{Bouchet(2011)},\cite{Hanany(2019)},\cite{Errard:2016}, 
might follow (or not)
an angular pattern similar to the temperature fluctuations seen in the sky, thereby shedding light on this tantalizing hypothesis.

\vskip 0.7cm

\subsection*{Acknowledgments}
I acknowledge inspiring discussions with V.Sanz
at the beginning of this work.
This work was partially funded by Spanish Agencia Estatal de Investigaci\'on
under grant PID2020-113334GB-I00 / AEI / 10.13039/501100011033, and 
by Generalitat
Valenciana under grant PROMETEO/2019/113 (EXPEDITE).

\end{document}